 \documentclass[prb,aps,preprint]{revtex4}

\usepackage{amsmath}
\usepackage{amssymb}
\begin{document}

\title{
Strong interaction of correlated electrons with phonons:
Exchange of phonon clouds by polarons
}
%
\author{V.\ A.\ Moskalenko\thanks{Permanent address:
Institute of Applied Physics, Moldova Academy of Sciences,
Chisinau 2028, Moldova}\cite{MosEmail}}
\affiliation{Joint Institute for Nuclear Research, 141--980 Dubna, Russia}
\author{P.\ Entel\cite{EnEmail}}
\affiliation{Theoretische Physik, Gerhard-Mercator-Universit\"at Duisburg,
        47048 Duisburg, Germany}
\author{M.\ Marinaro\cite{MaEmail}}
\affiliation{Dipartimento di Scienze Fisiche E.\ R.\ Caianiello,
        Universita degli Studi di Salerno, 84081 Baronissi, Italy}
\author{D.\ F.\ Digor\cite{DigEmail}}
\affiliation{Institute of Applied Physics, Moldova Academy of Sciences,
        Chisinau 2028, Moldova}
\date{\today}
%
%
\begin{abstract}
%
We investigate the interaction of strongly correlated electrons with
phonons in the frame of the Hubbard-Holstein model. The
electron-phonon interaction is considered to be strong and is an
important parameter of the model besides the Coulomb repulsion of
electrons and band filling. This interaction with the nondispersive
optical phonons has been transformed to the problem of mobile polarons
by using the canonical transformation of Lang and Firsov.
We discuss in particular the case for which the on-site Coulomb
repulsion is exactly cancelled by the phonon-mediated attractive
interaction and suggest that polarons exchanging phonon clouds
can lead to polaron pairing and superconductivity. It is then
the frequency of the collective mode of phonon clouds being
larger than the bare frequency, which determines the superconducting
transition temperature.
%
\end{abstract}

 \maketitle

\section{Introduction}

Since the discovery of high-temperature superconductivity by Bednorz
and M\"uller,\cite{Bednorz_86} the Hubbard model and related models
such as RVB and $t$-$J$ have widely been used to discuss the physical
properties of the normal and superconducting state.
\cite{Dagotto_94,Kampf_94,Scalapino_95,Brenig_95,Shen_95}
However, a unanimous explanation of the origin of the condensate in
high-temperature superconductors has not emerged so far.
One of the unsolved questions is in how far can phonons be
involved in the formation of the superconducting state.
In experimental and theoretical works mostly the change of phonon frequencies
and phonon life times associated with the superconducting transition
were discussed. For example, the decrease of frequencies of
Raman-active phonons at the transition \cite{Thomson_89} and
observation of the isotope effect for the case of not optimally
doped superconductors \cite{Frank_93}, as well as the observation of
phonon-induced structure in the tunnel characteristics,
\cite{Vedneev_92} speak in favor of strong electron-phonon coupling in
the cuprates.

The aim of the present paper is to gain further insight into the
mutual influence of strong on-site Coulomb repulsion and
strong electron-phonon interaction by using the single-band
Hubbard-Holstein model and a recently developed diagrammatic
approach.
\cite{Vladimir_90,Vakaru_90,Moskalenko_94,Bogoliubov_91,Bogoliubov_92}
For simplicity we consider coupling to dispersionless phonons only,
although this  might not be the most interesting case with
respect to superconductivity. However, previous investigations
\cite{Moskalenko_97a,Moskalenko_97b,Moskalenko_97c}
have shown that the Hubbard-Holstein model
\cite{Hubbard_63,Holstein_59} constitutes a formidable problem
of its own. Other authors have also intensively studied this
model Hamiltonian.
\cite{Freericks_94,Alexandrov_96,Alexandrov_00,Mierzejewski_98}

Because the interactions between electrons and electrons and
phonons are strong, we include the Coulomb repulsion in the
zero-order Hamiltonian and apply the canonical transformation of Lang
and Firsov \cite{Lang_63} in order to eliminate the linear
electron-phonon interaction. In the strong electron-phonon coupling
limit the resulting Hamiltonian of hopping polarons (i.e., hopping
electrons surrounded by clouds of phonons) can lead to an
attractive interaction among electrons being mediated by the
phonons. In this limit the chemical potential, on-site Coulomb
energy as well as the frequency of the collective mode of phonon clouds
(which is much larger than the bare frequency of the Einstein
oscillators) are strongly renormalized
\cite{Moskalenko_97c,Moskalenko_99a,Moskalenko_99b}
affecting the dynamical properties of the polarons and the
character of the superconducting transition. This will be
discussed by assuming that renormalized on-site Coulomb
repulsion and attractive electron-electron interaction completely
cancel each other. We suggest that the resulting
superconducting state with polaronic
Cooper pairs is mediated by the exchange of phonon
clouds during the hopping processes of the electrons.

\section{Theoretical approach}

\subsection{The Lang-Firsov transformation of the Hubbard-Holstein
            model}

The initial Hamiltonian of correlated electrons coupled to
optical phonons with bare frequency $\omega_0$ is given by
%
\begin{eqnarray}
%
{\cal H} & = & {\cal H}_e + {\cal H}_{ph}^0 + {\cal H}_{e-ph},
\\*[0.2cm]
{\cal H}_e & = & \sum\limits_{i j \sigma}
    \left \lbrace t(j-i) - \epsilon_0 \, \delta_{ij} \right \rbrace
    a_{j \sigma}^{\, \dagger} a_{i \sigma}
    + U \sum\limits_i n_{i_{\uparrow}} n_{i_{\downarrow}} ,
\\
{\cal H}_{ph}^0 & = & \sum\limits_i \hbar \omega_0 \,
    \left ( b_i^{\, \dagger} b_i + \textstyle{\frac{1}{2}} \right ) ,
    \qquad
{\cal H}_{e-ph}  =  g \sum\limits_i n_i q_i ,
\\
n_i & = & \sum\limits_\sigma n_{i\sigma} ,
\quad
n_{i \sigma} = a_{i \sigma}^{\dagger} a_{i \sigma} ,
\quad
q_i = \frac{1}{\sqrt{2}} \left ( b_i + b_i^{\, \dagger} \right ) .
%
\end{eqnarray}
%
Here $a_{i \sigma}^{\, \dagger}$ ($a_{i \sigma}$) and
$b_i^{\, \dagger}$ ($b_i$) are creation (annihilation) operators of
electrons and phonons, respectively; $i$ refers to the lattice site
and $\sigma$ to the spin; $q_i$ is the phonon coordinate and $g$ the
electron-phonon interaction constant;
$\epsilon_0 = \Bar{\epsilon}_0 - \mu$ with local energy
$\Bar{\epsilon}_0$ and chemical potential $\mu$; $U$ the on-site
Coulomb repulsion; $t(j-i)$ is the two-center transfer integral. The
Fourier representation of $t(j-i)$ is connected to the tight-binding
dispersion $\varepsilon ({\bf k})$ of the bare electrons,
$$
t(j-i) = \frac{1}{N} \sum\limits_{\bf k} \varepsilon ({\bf k})
         \exp \{-i {\bf k} ({\bf R}_j - {\bf R}_i) \} ,
$$
with band width $W$.
The energy scale of this model is fixed by the parameters $W$, $U$,
$g$ and $\hbar \omega_0$. As an additional parameter we have the band
filling.

After applying the Lang-Firsov transformation \cite{Lang_63}
%
\begin{equation}
{\cal H}_p = e^S {\cal H} e^{-S}, \quad
c_{i \sigma} = e^S a_{i \sigma} e^{-S}, \quad
c_{i \sigma}^{\dagger} = e^S a_{i\sigma}e^{-S}
%
\end{equation}
%
with
%
\begin{equation}
%
S = - i \Bar{g} \sum\limits_i n_i p_i, \quad
\Bar{g} = \frac{g}{\hbar \omega_0},    \quad
p_i = \frac{i}{\sqrt{2}} \left( b_i^{\, \dagger} - b_i \right) ,
%
\end{equation}
%
where $p_i$ is the phonon momentum and $\Bar{g}$ the
dimensionless interaction constant, the polaron Hamiltonian is
obtained as
%
\begin{eqnarray}
%
{\cal H}_p & = &{\cal H}_p^0 + {\cal H}_{ph}^0
                   + {\cal H}_{int} ,
\\*[0.2cm]
{\cal H}_p^0 & = & \sum\limits_i {\cal H}_{ip}^0,
  \quad {\cal H}_{ip}^0 = \epsilon \sum\limits_\sigma
  n_{i \sigma} + \Bar{U} n_{i \uparrow} n_{i \downarrow},
\\
{\cal H}_{int} & = & \sum\limits_{i,j,\sigma} t(j-i)
     c_{j \sigma}^{\, \dagger} c_{i \sigma},
\\
c_{i\sigma}^{\, \dagger} & = &
      a_{i \sigma}^{\, \dagger} e^{-i \Bar{g} p_i}, \quad
c_{i \sigma} =
     a_{i \sigma} e^{i \Bar{g} p_i},
\\
\epsilon & = & \Bar{\epsilon}_0 - \Bar{\mu}, \quad
  \Bar{\mu} = \mu + \alpha \hbar \omega_0, \quad
  \Bar{U} = U - 2 \alpha \hbar \omega_0, \quad
  \alpha = \textstyle{\frac{1}{2}} \Bar{g}^2 .
%
\end{eqnarray}
%
In order to derive the polaron Hamiltonian it was necessary to
include the shift of the phonon coordinate $q_{_i}$ of the form
$e^S q_i e^{-S} = q_i - \Bar{g} n_i$ which is responsible for
the elimination of the linear electron-phonon interaction.
The polaron Hamiltonian is by nature a polaron-phonon operator,
i.e., the creation operator $c_{i \sigma}^{\dagger}$
and destruction operator $c_{i \sigma}$ in ${\cal H}_p$
should be interpreted as creation and
destruction operators of polarons (electrons dressed with
the displacements of the ions) which couple dynamically to the
momentum of the optical phonon. In zero-order approximation
(omitting ${\cal H}_{int}$) polarons and phonons are localized
with strongly renormalized chemical potential $\Bar{\mu}$ and on-site
Coulomb interaction $\Bar{U}$. The operator
${\cal H}_{int}$ describes the tunneling of polarons between
the lattice sites, i.e., tunneling of electrons surrounded by
clouds of phonons.

\subsection{Expansion about the atomic limit}

The problem is now to deal properly with the impact of electronic
correlations on the polaron problem. This can be done best by using
Green's functions provided one finds a key to deal with
the spin and charge degrees of freedom.
For the general case when $\Bar{U}$ is different from zero, the
Coulomb interaction has to be included in the zero-order
Hamiltonian. As a consequence conventional perturbation theory of
quantum statistical mechanics is not an adequate tool because
it relies on the expansion of the partition function about the
noninteracting state (use of traditional Wick's theorem and
conventional Feynman diagrams). Similar is the situation for
composite particles like polarons,
$c_{i\sigma} = a_{i\sigma} \exp ({i\Bar{g}p_i})$, involving operators
for the electron and phonon subsystem.

It was Hubbard \cite{Hubbard_67} who proposed a graphical expansion for
correlated electrons about the atomic limit in powers of the hopping
integrals. This diagrammatic approach was reformulated in a systematic way
for the single-band Hubbard model by Slobodyan and Stasyuk,
\cite{Slobodyan_74} and, independently, by Zaitsev\cite{Zaitsev_76}
and further developed by Izyumov. \cite{Izyumov_90} In these approaches
the complicated algebraical structure of projection or Hubbard
operators was used. Therefore, it appeared to be more appropriate to
develop a diagrammatic technique which uses more simple creation and
annihilation operators for electrons at all intermediate stages of the
theory (for details see Ref.\ \onlinecite{Vladimir_90,Vakaru_90}).
In the latter approach the averages of chronological products of
interactions are reduced to $n$-particle Matsubara Green's functions
of the atomic system. These functions can be factorized into
independent local averages by using a generalization of Wick's
theorem (GWT), which takes into account strong local correlations
(details are given in Ref.\
\onlinecite{Vladimir_90,Vakaru_90,Moskalenko_99a}).
The application of GWT yields new irreducible on-site many-particle
Green's functions or Kubo cumulants. These new functions contain all
local spin and charge fluctuations. The analogical linked-cluster
expansion for the Hubbard model around the atomic limit
was recently reformulated by Metzner. \cite{Metzner_91}

\subsection{Averages of phonon operators}

We define the temperature Green's function for the polarons in (7)
in the interaction representation by
%
\begin{equation}
%
{\cal G}({\bf x}, \sigma , \tau |
         {\bf x}^{\prime}, \sigma^{\prime}, \tau^{\prime})
= - \langle T \, c_{{\bf x} \sigma}(\tau) \,
    \Bar{c}_{{\bf x}^{\prime} \sigma^{\prime}}
    (\tau^{\prime}) \, U(\beta) \rangle_{\! 0}^{\! c} ,
%
\end{equation}
%
with
$$
c_{{\bf x} \sigma }(\tau)
  = e^{{\cal H}^0 \tau} c_{{\bf x} \sigma} e^{-{\cal H}^0 \tau} , \quad
\Bar{c}_{{\bf x} \sigma}(\tau)
  = e^{{\cal H}^0 \tau} c_{{\bf x} \sigma}^{\, \dagger}
    e^{-{\cal H}^0 \tau} ,
$$
with ${\cal H}^0 = {\cal H}_p^0 + {\cal H}_{ph}^0$ and evolution
operator given by
%
\begin{equation}
%
U(\beta) = T \exp \left (
         - \int_0^{\beta} \! d\tau \, H_{int}(\tau)
           \right ) ,
%
\end{equation}
%
and where $\bf x, x^{\prime}$ are the site indices and
$\tau,\tau^{\prime}$ stand for the imaginary time with
$0 < \tau < \beta$; $T$ is the time ordering operator and
$\beta$ the inverse temperature. The statistical average
$\langle \ldots \rangle_0^c$ is evaluated with respect to
the zero-order density matrix of the grand-canonical ensemble
of the localized polarons and phonons
%
\begin{equation}
%
\frac{e^{- \beta {\cal H}^0}}
      {\operatorname{Tr} e^{- \beta {\cal H}^0}}
  = \prod_i
\frac{e^{- \beta {\cal H}_{i \, p}^0}}
      {\operatorname{Tr} e^{- \beta {\cal H}_{i \, p}^0}}
\frac{e^{- \beta {\cal H}_{i \, ph}^0}}
      {\operatorname{Tr} e^{- \beta {\cal H}_{i \, ph}^0}}
%
\end{equation}
%
The upper index $`c'$ in (13) means that only connected diagrams
must be taken into account. The density matrix (14) is factorized
with respect to the lattice sites. The phonon part is easily
diagonalized by using the free phonon operators, $b_i$ and
$b_i^{\dagger}$, while the on-site polaron Hamiltonian contains
the polaron-polaron interaction which is proportional to the
renormalized parameter $\Bar{U}$, which only can be diagonalized
by using Hubbard operators. \cite{Hubbard_63}
At this stage no special assumption is made about the quantity
$\Bar{U}$ and its sign and we will set up the equations of motion
for the dynamical quantities for this general case. However,
a detailed inspection of the equations will only be undertaken for
the special case $\Bar{U} = 0$.

Wick's theorem of weak-coupling quantum field theory can be used
when evaluating statistical averages of phonon operators like,
for example, the propagator for the phonon cloud,
%
\begin{eqnarray}
%
\Phi(\tau_1|\tau_2) =
\Phi (\tau_1 - \tau_2) & \equiv &
      \langle
      T \exp \{ i \Bar{g} \, [ p(\tau_1) - p(\tau_2) ] \}
      \rangle_0
\nonumber \\
& = & \exp \left( -\textstyle{\frac{1}{2}} \Bar{g}^2
      \langle T \, \left[ p(\tau_1) - p(\tau_2) \right]^2
      \rangle_0 \right)
\nonumber \\
& = & \exp \left( - \sigma(\beta) + \sigma(|\tau_1 - \tau_2|)
      \right) ,
%
\end{eqnarray}
%
\begin{eqnarray}
%
\Phi(\tau_1,\tau_2|\tau_3,\tau_4) \equiv
\langle \! & \! T \! & \exp \{ i\Bar{g} \,
       [ p(\tau_1) + p(\tau_2) - p(\tau_3) - p(\tau_4) ] \}
       \rangle_0 =
\nonumber \\
& = &
    \exp \left(
  - \textstyle{\frac{1}{2}} \Bar{g}^2
    \langle T \, [p(\tau_1) + p(\tau_2) - p(\tau_3)
  - p(\tau_4)]^2 \rangle_0 \right)
\nonumber \\
& = & \exp \{ \sigma (|\tau_1 - \tau_3|)
  + \sigma (|\tau_1 - \tau_4|)
  + \sigma (|\tau_2 - \tau_3|)
\nonumber \\
& + & \sigma (|\tau_2 - \tau_4|)
  -   \sigma (|\tau_1 - \tau_2|) -\sigma (|\tau_3-\tau_4|)
\nonumber \\
& - & 2 \sigma (\beta) \} ,
%
\end{eqnarray}
%
where
%
\begin{eqnarray}
%
\sigma \left( |\tau_1 - \tau_2| \right)
& = & \Bar{g}^2 \langle T \,
      p(\tau_1) p(\tau_2) \rangle_0
\nonumber \\*[0.2cm]
& = & \alpha \, \frac{
      \displaystyle \cosh \left( \hbar \omega_0 \left \lbrace
      \frac{\beta}{2} - |\tau_1 - \tau_2| \right \rbrace
      \right) }
      {\displaystyle \sinh \left(
      \frac{\beta \hbar \omega_0}{2} \right) } .
%
\end{eqnarray}
%

We now have to discuss the problem of how to calculate
chronological averages of combinations of polaron operators.
Here we will make use of the above mentioned new
diagram technique and the GWT.\cite{Vladimir_90,Vakaru_90}
This approach has many-particle on-site irreducible Green's
functions as main element of the diagrams.

\section{Polaron and phonon Green's functions}

In zero-order approximation the one-polaron Green's function
is of the form
%
\begin{eqnarray}
%
{\cal G}_p^0(x | x^{\prime})
& = & - \langle T \,
     {c}_{{\bf x} \sigma}(\tau)
     \Bar{c}_{{\bf x}^{\prime} \sigma^{\prime}}
     (\tau^{\prime}) \rangle_0
\nonumber \\
& = & - \langle T \,
     a_{{\bf x} \sigma}(\tau)
     \Bar{a}_{{\bf x}^{\prime} \sigma^{\prime}}
     (\tau^{\prime}) \rangle_0 \Phi(\tau|\tau^{\prime})
\nonumber \\
& = & {\cal G}^{(0)} (x | x^{\prime})
      \, \Phi(\tau|\tau^{\prime}) ,
%
\end{eqnarray}
%
with $x=({\bf x}, \sigma, \tau)$.
The simplest new element of the diagram technique is the
two-particle irreducible Green's function or Kubo cumulant,
which is equal to
%
\begin{alignat}{2}
%
{\cal G}_2^{(0)\, ir} (x_1,x_2 | x_3,x_4)
 = &\;\, \delta_{{\bf x}_1, {\bf x}_2} \delta_{{\bf x}_1, {\bf x}_3}
     \delta_{{\bf x}_1, {\bf x}_4}
\nonumber \\
& \times
{\cal G}_2^{(0)\, ir} (\sigma_1, \tau_1; \sigma_2, \tau_2 |
                       \sigma_3, \tau_3; \sigma_4, \tau_4) ,
%
\end{alignat}
%
where
%
\begin{alignat}{4}
%
{\cal G}_2^{(0)\, ir} & (\sigma_1, \tau_1; \sigma_2, \tau_2 |
                           \sigma_3, \tau_3; \sigma_4, \tau_4)
\nonumber \\
 =  &\;\, \langle T \,
      c_{\sigma_1}(\tau_1) \, c_{\sigma_2}(\tau_2)
      \Bar{c}_{\sigma_3}(\tau_3) \, \Bar{c}_{\sigma_4}(\tau_4) \rangle_0
\nonumber \\
& - \; \langle T \,
      c_{\sigma_1}(\tau_1) \, \Bar{c}_{\sigma_4}(\tau_4) \rangle_0
      \langle T \,
      c_{\sigma_2}(\tau_2) \, \Bar{c}_{\sigma_3}(\tau_3) \rangle_0
\nonumber \\
& +  \; \langle T \,
      c_{\sigma_1}(\tau_1) \, \Bar{c}_{\sigma_3}(\tau_3) \rangle_0
      \langle T \,
      c_{\sigma_ 2}(\tau_2) \, \Bar{c}_{\sigma_4}(\tau_4) \rangle_0 .
%
\end{alignat}
%
The first term on the right-hand side of Eq.\ (21) is of the
form
%
\begin{alignat}{2}
%
\langle & T \,
   c_{\sigma_1}(\tau_1) \, c_{\sigma_2}(\tau_2) \,
   \Bar{c}_{\sigma_3}(\tau_3) \, \Bar{c}_{\sigma_4}(\tau_4) \rangle_0
\nonumber \\
 & =  \langle T \,
   a_{\sigma_1}(\tau_1) \, a_{\sigma_2}(\tau_2) \,
   \Bar{a}_{\sigma_3}(\tau_3) \, \Bar{a}_{\sigma_4}(\tau_4) \rangle_0
   \, \Phi(\tau_1 \tau_2|\tau_3 \tau_4)
%
\end{alignat}
%

When the number of polaron operators increases more complicated
irreducible Green's Functions like
${\cal G}_n^{(0) ir} (x_1 \ldots x_n|x_1^{\prime} \ldots
   x_n^{\prime})$
with $n \geq 3$ and all possible terms of their products will
appear. The sum of all {\em strongly connected diagrams} (i.e.,
those which cannot be divided into two parts by cutting a single
hopping line) containing all kinds of irreducible Green's functions
in the perturbation expansion of the evolution
operator, defines the special function $Z(x|x^{\prime})$
(for details see Ref.\ \onlinecite{Vladimir_90,Vakaru_90}).
This function contains all contributions from charge and spin
fluctuations. It allows us, together with the mass operator which
is in our case the hopping matrix element, to formulate a
Dyson-type of equation for the one-polaron Green's function,
\cite{Vladimir_90,Vakaru_90,Moskalenko_94,Bogoliubov_91,Bogoliubov_92}
%
\begin{equation}
%
{\cal G}(x|x^{\prime})
= \Lambda (x|x^{\prime})
+ \sum\limits_{1,2} \Lambda (x|1) t(1-2)
  {\cal G}(2|x^{\prime}) ,
%
\end{equation}
%
where
%
\begin{eqnarray}
%
\Lambda (x|x^{\prime})
& = & {\cal G}_p^{(0)}(x|x^{\prime})
      + Z(x|x^{\prime}) ,
\\
t(x - x^{\prime})
& = & \delta _{\sigma, \sigma^{\prime}}
      \delta (\tau - \tau^{\prime}) \,
      t({\bf x} - {\bf x}^{\prime}) .
%
\end{eqnarray}
%
Here $x$ stands again for ${\bf x}, \sigma, \tau$ and the sum
is over the discrete indices including the integration
over $\tau $.
Using the Fourier representation for these quantities,
%
\begin{eqnarray}
%
{\cal G}({\bf x}|\tau)
 & = & \frac{1}{N} \sum_{\bf k} \frac{1}{\beta} \sum_{\omega_n}
       e^{-i{\bf kx} -i\omega_n \tau}
       {\cal G}_{\sigma}({\bf k}|i\omega_n) ,
\nonumber \\
\Lambda({\bf x}|\tau)
 & = & \frac{1}{N} \sum_{\bf k} \frac{1}{\beta} \sum_{\omega_n}
       e^{-i{\bf kx} -i\omega_n \tau}
       \Lambda_{\sigma}({\bf k}|i\omega_n) ,
\nonumber \\
{\cal G}({\bf x}|i\omega_n)
 & = & \frac{1}{2} \int_{-\beta}^{+\beta} \!\! d\tau \, e^{i\omega_n \tau}
       {\cal G}_{\sigma}({\bf x}|\tau) ,
\nonumber \\
\Lambda({\bf x}|i\omega_n)
 & = & \frac{1}{2} \int_{-\beta}^{+\beta} \!\! d\tau \, e^{i\omega_n \tau}
       \Lambda_{\sigma}({\bf x}|\tau) ,
%
\end{eqnarray}
%
gives us the following form of the Dyson equation for the renormalized
one-polaron Green's function,
%
\begin{equation}
%
{\cal G}_\sigma ({\bf k} | i\omega_n)
  = \frac{\Lambda_{\sigma}({\bf k})}
         {1 - \varepsilon({\bf k}) \, \Lambda({\bf k} | i\omega_n)}
%
\end{equation}
%
Here
$\omega_n$ stands for the odd Matsubara frequency,
$\omega_n = (2n + 1) \, \pi/\beta$.

For further discussion of ${\cal G}_{\sigma}({\bf} k | i \omega)$
we need the Fourier representation of the zero-order
one-polaron Green's function ${\cal G}_p^{(0)}$ defined
in (18). In order to facilitate the investigation we have
evaluated the propagator of the phonon cloud (16) in the
strong-coupling limit $\alpha \gg 1$,
\cite{Moskalenko_97a,Moskalenko_97b,Moskalenko_99b}
%
\begin{eqnarray}
%
\Phi (\tau) & = &
   \frac{1}{\beta} \sum\limits_{\Omega _n}
   e^{-i \Omega_n(\tau)}
   \Bar{\Phi}(i \Omega_n) ,
\\
\Bar{\Phi}(i \Omega_n) & = &
   \frac{e^{-\sigma (\beta)}}{2} \int_{-\beta }^\beta
   \!\! d\tau \, e^{\, i \Omega_n \tau + \sigma (|\tau |)}
%
\end{eqnarray}
%
with $\Omega_n = 2n\pi/\beta$.
In order to find $\Bar{\Phi}(i \Omega_n)$ we use the Laplace
approximation \cite{Fedoryuk_87} for the integral (28) which contains
an exponential function with the parameter $\alpha$. In the
strong-coupling limit $\alpha \gg 1$ we have obtained
%
\begin{eqnarray}
%
\Bar{\Phi}(i \Omega n) \approx
    \frac{2 \omega_c}{\Omega_n^2 + \omega_c^2} ,
\quad
\omega_c =  \hbar \alpha \omega_0 = \hbar \frac{g^2}{2\hbar \omega_0} .
%
\end{eqnarray}
%
This term is the harmonic propagator of the
collective mode of phonons belonging to the polaron clouds.
There are further terms describing anharmonic deviations.
For $\alpha \gg 1$ these terms will be omitted since they are
smaller compared to the harmonic contribution. We obtain then
for the Fourier representation of the phonon correlation
function by making use of the Laplace approximation
\cite{Fedoryuk_87} and
%
\begin{align}
%
\Phi(\tau_1,\tau_2|\tau_3,\tau_4) &=  \frac{1}{\beta^4}
  \sum\limits_{\Omega_1 \ldots \Omega_4} \!\!
  \Bar{\Phi}(i\Omega_1,i\Omega_2|i\Omega_3,i\Omega_4) \,
  e^{- i\Omega_1\tau_1 - i\Omega_2\tau_2
     + i\Omega_3\tau_3 + i\Omega_4\tau_4} ,
\\
\Bar{\Phi}(i\Omega_1,i\Omega_2|i\Omega_3,i\Omega_4) &=
  \int_0^{\beta} \ldots \int_0^{\beta} \! \! d\tau_1 \ldots d\tau_4 \,
  e^{i\Omega_1\tau_1 + i\Omega_2\tau_2
    - i\Omega_3\tau_3 - i\Omega_4\tau_4} ,
\\
\intertext{the result}
\Bar{\Phi}(i\Omega_1,i\Omega_2|i\Omega_3,i\Omega_4) &\simeq
  \left [
  \delta_{\Omega_1,\Omega_3} \delta_{\Omega_2,\Omega_4}
+ \delta_{\Omega_1,\Omega_4} \delta_{\Omega_2,\Omega_3}
  \right ] \, \Bar{\Phi}(i\Omega_1) \, \Bar{\Phi}(i\Omega_2) ,
\\
\intertext{which corresponds to}
\Phi(\tau_1,\tau_2|\tau_3,\tau_4) &\simeq
  \Phi(\tau_1|\tau_3) \, \Phi(\tau_2|\tau_4)
+ \Phi(\tau_1|\tau_4) \, \Phi(\tau_2|\tau_3) .
%
\end{align}
%
This means that we keep in the following only the free collective
oscillations of the phonon clouds (29) which surround the polarons
and use the Hartree-Fock approximation (32) and (33) for their
two-particle correlation functions. In particular we will investigate
the influence of absorption and emission of this collective mode
(by the polarons) on the superconducting phase transition.

By using the harmonic mode (29) the Fourier representation of the
local polaron Green's function
%
\begin{equation}
%
\Bar{\cal G}_{p\sigma}^{(0)}(i\omega_n)
= \frac{1}{2} \int_{-\beta}^\beta \! \! d\tau \,
  e^{i \omega_n \tau} \Bar{\cal G}_{p\sigma}^{(0)}(\tau)
%
\end{equation}
%
is equal to
%
\begin{align}
%
\Bar{\cal G}_{p\sigma}^{(0)}(i\omega_n) \simeq
  \frac{1}{Z_0} & \left(
  \frac{e^{-\beta E_0}
        + \Bar{N}(\omega_c)(e^{-\beta E_0} + e^{-\beta E_\sigma})}
       {i\omega_n + E_0 - E_{\sigma} - \omega_c}
       \right .
\nonumber \\*[0.2cm]
& + \frac{e^{-\beta E_\sigma}
        + \Bar{N}(\omega_c)(e^{-\beta E_0} + e^{-\beta E_\sigma})}
       {i\omega_n + E_0 - E_{\sigma} + \omega_c}
\nonumber \\*[0.2cm]
& + \frac{e^{-\beta E_{-\sigma }}
        + \Bar{N}(\omega_c)(e^{-\beta E_{\sigma}} + e^{-\beta E_2})}
       {i\omega_n +E_{-\sigma} - E_2 - \omega_c}
\nonumber \\*[0.2cm]
& + \left . \frac{e^{-\beta E_2}
        + \Bar{N}(\omega_c)(e^{-\beta E_{\sigma}} + e^{-\beta E_2})}
       {i\omega_n + E_{-\sigma} -E_2 + \omega_c}
  \right) ,
\label{g_approx}
%
\end{align}
%
where
%
\begin{subequations}\begin{align}
%
Z_0 & = 1
      + e^{-\beta E_{\sigma}} + e^{-\beta E_{-\sigma}} + e^{-\beta E_2} ,
 \\
E_0 & = 0, \quad E_{\pm \sigma} = \epsilon, \quad E_2 = \Bar{U} + 2\epsilon ,
%
\end{align}\end{subequations}
%
\begin{equation}
%
\Bar{n}(\epsilon) = \left ( e^{\beta\epsilon} + 1 \right )^{-1} ,
\quad \Bar{N}(\omega_c) = \left ( e^{\beta\omega_c} - 1 \right )^{-1}.
%
\end{equation}
%
Equation (35) shows that the on-site transition energies of
the polarons are changed by the collective-mode energy
$\pm \omega_c$ of the phonon clouds. The delocalization of
the polarons due to their hopping between the lattice sites causes
the broadening of the polaron energy levels. Equation (35) can be
further simplified for the case of small on-site interaction
energy $\Bar{U}$ of polarons. For $\Bar{U} = 0$ we obtain
%
\begin{align}
%
\Bar{\cal G}_{p\sigma}^{(0)}(i\omega_n|\epsilon) &=
  \frac{\Bar{N}(\omega_c) + 1 - \Bar{n}(\epsilon)}
  {i\omega_n - \epsilon - \omega_c}
+ \frac{\Bar{N}(\omega_c) + \Bar{n}(\epsilon)}
  {i\omega_n - \epsilon + \omega_c}
\nonumber \\*[0.2cm]
&= \frac{(i\omega_n - \epsilon) \coth (\beta\omega_c/2)
   + \omega_c \tanh (\beta\epsilon/2)}
   {(i\omega_n - \epsilon)^2 - \omega_c^2} .
%
\end{align}
%
This function has the antisymmetric property
%
\begin{equation}
%
\Bar{\cal G}_{p\sigma}^{(0)}(-i\omega_n|-\epsilon)
 = -  \Bar{\cal G}_{p\sigma}^{(0)}(i\omega_n|\epsilon)
%
\end{equation}
%
which holds also for the renormalized polaron quantities,
%
\begin{align}
%
\Lambda_{\sigma}(-{\bf k}, -i\omega_n|-\epsilon) &= -
  \Lambda_{\sigma}({\bf k}, i\omega_n|\epsilon) ,
  \nonumber \\
{\cal G}_{\sigma}(-{\bf k}, -i\omega_n|-\epsilon) &= -
  {\cal G}_{\sigma}({\bf k}, i\omega_n|\epsilon) .
%
\end{align}
%
When taking $\Bar{U} \simeq 0$ we assume that the strong on-site
Coulomb repulsion of polarons can be canceled by the attraction
induced by the strong electron-phonon interaction. We consider this as
a model case which allows to discuss in a transparent manner the
polarons exchanging phonon clouds during hopping between the lattice
sites.

\section{Two-particle irreducible correlation functions}

In the following we discuss the influence of strong electron-phonon
interaction on the two-particle irreducible Green's function. For
$\Bar{U} = 0$ the electronic correlation function in (22)
is equal to
%
\begin{align}
%
\langle T & \, a_{\sigma_1}(\tau_1) \, a_{\sigma_2}(\tau_2) \,
  \Bar{a}_{\sigma_3}(\tau_3) \, \Bar{a}_{\sigma_4}(\tau_4) \rangle_{0}
\nonumber \\
 = & \; \langle T \,
   a_{\sigma_1}(\tau_1) \Bar{a}_{\sigma_4}(\tau_4) \rangle_{0}
    \langle  T \, a_{\sigma_2}(\tau_2) \Bar{a}_{\sigma_3}(\tau_3) \rangle_{0}
\nonumber \\
& -  \langle T \, a_{\sigma_1}(\tau_1) \Bar{a}_{\sigma_3}(\tau_3) \rangle_{0}
    \langle T \, a_{\sigma_2}(\tau_2) \Bar{a}_{\sigma_4}(\tau_4) \rangle_{0}
%
\end{align}
%
because usual Wick's theorem works now. Making use of (33) we obtain
for the two-particle irreducible Green's function (21) the relation,
%
\begin{align}
%
{\cal G}_2^{(0) \, ir} & (\sigma_1,\tau_1;\sigma_2,\tau_2|
     \sigma_3,\tau_3;\sigma_4\tau_4)
\nonumber \\
= & \; \delta_{\sigma_1,\sigma_4} \delta_{\sigma_2,\sigma_3} \,
  {\cal G}_{\sigma_1}^{(0)}(\tau_1 - \tau_4) \,
  {\cal G}_{\sigma_2}^{(0)}(\tau_2 - \tau_3) \,
  \Phi(\tau_1 - \tau_3) \, \Phi(\tau_2 - \tau_4)
\nonumber \\
& - \delta_{\sigma_1,\sigma_3} \delta_{\sigma_2,\sigma_4} \,
  {\cal G}_{\sigma_1}^{(0)}(\tau_1 - \tau_3) \,
  {\cal G}_{\sigma_2}^{(0)}(\tau_2 - \tau_4) \,
  \Phi(\tau_1 - \tau_4) \, \Phi(\tau_2 - \tau_3) .
%
\end{align}
%
In the absence of exchange of phonon clouds by the polarons this
quantity has to be zero. Indeed, if during the time of propagation
of two polarons the electrons keep their initial phonon clouds, then
the irreducible two-polaron Green's function (21) will vanish for
the case $\Bar{U} = 0$. But because two electrons can be exchanged
(independently of the exchange of phonon clouds) we obtain new
contributions corresponding to two polarons with exchanged phonon
clouds. Alternatively we may say that, in the case $\Bar{U} = 0$,
Wick's theorem works separately for free electrons and free phonons;
however, it does not apply for polarons as composite particles.
Hence their cumulants do not vanish.

The Fourier representation of (42),
%
\begin{align}
%
{\cal G}_2^{(0) \, ir} & (\sigma_1,i\omega_1;\sigma_2,i\omega_2|
                          \sigma_3,i\omega_3;\sigma_4,i\omega_4)
\nonumber \\
= & \; \int_0^{\beta} \ldots \int_0^{\beta}
       \!\! d\tau_1 \ldots d\tau_4 \,
       {\cal G}_2^{(0) \, ir} (\sigma_1,\tau_1;\sigma_2,\tau_2|
                               \sigma_3,\tau_3;\sigma_4,\tau_4) \,
       e^{i\omega_1\tau_1 + i\omega_2\tau_2 - i\omega_3\tau_3
        - i\omega_4\tau_4} ,
%
\end{align}
%
has the form
%
\begin{align}
%
{\cal G}_2^{(0) \, ir} & (\sigma_1,i\omega_1;\sigma_2,i\omega_2|
                          \sigma_3,i\omega_3;\sigma_4,i\omega_4)
\nonumber \\ = & \; \beta \,
          \delta_{\omega_1 + \omega_2, \omega_3 + \omega_4} \,
\Bar{\cal G}_2^{(0) \, ir} (\sigma_1,i\omega_1;\sigma_2,i\omega_2|
                          \sigma_3,i\omega_3;\sigma_4,i\omega_4)
\nonumber \\ = & \; \beta \,
          \delta_{\omega_1 + \omega_2, \omega_3 + \omega_4} \,
\left\lbrace
   \delta_{\sigma_1,\sigma_4} \, \delta_{\sigma_2,\sigma_3} \,
   A_{\sigma_1\sigma_2} (\sigma_1,i\omega_1;\sigma_2,i\omega_2|
                          \sigma_2,i\omega_3;\sigma_1,i\omega_4)
   \right .
\nonumber \\ & - \left .
   \delta_{\sigma_1,\sigma_3} \, \delta_{\sigma_2,\sigma_4} \,
   A_{\sigma_1\sigma_2} (\sigma_1,i\omega_1;\sigma_2,i\omega_2|
                          \sigma_1,i\omega_4;\sigma_2,i\omega_3)
\right\rbrace ,
%
\end{align}
%
where
%
\begin{align}
%
A_{\sigma_1\sigma_2} &(\sigma_1,i\omega_1;\sigma_2,i\omega_2|
                      \sigma_2,i\omega_3;\sigma_1,i\omega_4)
\nonumber \\ &= \frac{1}{\beta} \sum_{\Omega}
  \frac{(2 \omega_c)^2}
  {[i\omega_1 - \Omega - \epsilon][i\omega_3 - \Omega - \epsilon]
   [\Omega^2 + \omega_c^2][(\Omega + \Omega_1)^2 + \omega_c^2]}
%
\end{align}
%
with $\Omega_1 = \omega_2 - \omega_3$. The summation leads to
%
\begin{align}
%
A_{\sigma_1\sigma_2} &(\sigma_1,i\omega_1;\sigma_2,i\omega_2|
                      \sigma_2,i\omega_3;\sigma_1,i\omega_4)
                      = 2 (\omega_c)^2
\nonumber \\*[0.2cm] &\times \left \lbrace
\frac{\tanh(\beta\epsilon/2) \,
      [i\omega_1 + i\omega_2 - 2\epsilon]
      [2\omega_c^2 - (i\omega_1 - \epsilon)(i\omega_4 - \epsilon)
                   - (i\omega_2 - \epsilon)(i\omega_3 - \epsilon)]}
     {[(i\omega_1 - \epsilon)^2 - \omega_c^2]
      [(i\omega_2 - \epsilon)^2 - \omega_c^2]
      [(i\omega_3 - \epsilon)^2 - \omega_c^2]
      [(i\omega_4 - \epsilon)^2 - \omega_c^2]} \right .
\nonumber \\*[0.2cm] &-
\frac{2\omega_c[i\omega_1 + i\omega_2 -2\epsilon]^2
      \coth(\beta\omega_c/2) \,
      [2\omega_c^2 - (i\omega_1 - \epsilon)(i\omega_3 - \epsilon)
                   - (i\omega_2 - \epsilon)(i\omega_4 - \epsilon)]}
     {[(i\Omega_1)^2 - (2\omega_c)^2]
      [(i\omega_1 - \epsilon)^2 - \omega_c^2]
      [(i\omega_2 - \epsilon)^2 - \omega_c^2]
      [(i\omega_3 - \epsilon)^2 - \omega_c^2]
      [(i\omega_4 - \epsilon)^2 - \omega_c^2]}
\nonumber \\*[0.2cm] &-
\frac{\coth(\beta\omega_c/2)}
     {\omega_c [(i\Omega_1)^2 - (2\omega_c)^2]} \left [
\frac{(i\omega_1 - \epsilon)(i\omega_3 - \epsilon) + 3 \omega_c^2}
     {[(i\omega_1 - \epsilon)^2 - \omega_c^2]
      [(i\omega_3 - \epsilon)^2 - \omega_c^2]} \right .
\nonumber \\*[0.2cm] &+ \left . \left .
\frac{(i\omega_2 - \epsilon)(i\omega_4 - \epsilon) + 3 \omega_c^2}
     {[(i\omega_2 - \epsilon)^2 - \omega_c^2]
      [(i\omega_4 - \epsilon)^2 - \omega_c^2]} \right ] \right \rbrace
%
\end{align}
%
The function $A_{\sigma_1\sigma_2}$ contains the contributions
to the two-particles on-site Green's function from the different
spin channels. The spin structure in Eq.\ (44) is due to the
conversation law for the spins of the polarons.

\section{Superconducting phase transition}

In the following we check whether the polaronic system may
have a superconducting instability in the absence of a direct
attractive interaction for the polarons, i.e., for $\Bar{U} = 0$.
In this case the attraction is only brought about dynamically by
polarons exchanging phonon clouds. With respect to superconductivity
we need in addition to the normal state Green's function (13) the
anomalous propagators. \cite{Abrikosov_62} For simplicity we limit
the discussion to $s$-wave superconductivity as in previous
investigations of superconducting instabilities in the Hubbard
model \cite{Bogoliubov_91,Bogoliubov_92} and Hubbard-Holstein model
in the strong-coupling limit, $\alpha \gg 1$.
\cite{Moskalenko_99b}

For a description of the superconducting state we need the three
irreducible functions $\Lambda_{\sigma}$, $Y_{\sigma,-\sigma}$
and $\Bar{Y}_{-\sigma,\sigma}$ which represent infinite sums of
diagrams containing irreducible many-particle Green's functions.
In order to obtain a close set of equations we will restrict
ourselves to a class of rather simple contributions which,
however, contain the most important charge, spin and pairing
correlations; for details see Ref.\ \onlinecite{Moskalenko_99b}.
This class of diagrams is obtained by neglecting contributions
for which the Fourier representation of the superconducting order
parameters, $Y_{\sigma,-\sigma}$ and $\Bar{Y}_{-\sigma,\sigma}$,
depend on the polaron momentum ${\bf k}$. In this approximation
$Y_{\sigma,-\sigma}$ has to be obtained from
%
\begin{align}
%
Y_{\sigma,-\sigma}(i\omega) = &- \frac{1}{\beta N} \sum_{{\bf k},\omega_l}
  \frac{\varepsilon({\bf k}) \, \varepsilon(-{\bf k})
        Y_{\sigma,-\sigma}(i\omega_l)}
       {D_{\sigma}({\bf k}, i\omega_l)}
\nonumber \\*[0.2cm] &\times
  \Bar{\cal G}_2^{(0) \, ir}
  (\sigma,i\omega;-\sigma,-i\omega|\sigma,i\omega_l;-\sigma,-i\omega_l) .
%
\end{align}
%
In the same approximation $\Lambda_{\sigma}$ has to be computed
from
%
\begin{align}
%
\Lambda_{\sigma}(i\omega) = & \;{\cal G}_{p\sigma}^{(0)}(i\omega)
\nonumber \\ &- \frac{1}{\beta N} \sum_{{\bf k},\omega_l}
      \frac{\varepsilon^2({\bf k})}
           {D_{\sigma}({\bf k},i\omega_l)} \lbrace
      \Lambda_{\sigma}(i\omega_l) [
      1- \varepsilon(-{\bf k}) \,
      \Lambda_{-\sigma}(-i\omega_l) ]
\nonumber \\ &-
  \varepsilon({\bf k})\,
  Y_{\sigma,-\sigma}(i\omega_l) \,
  \Bar{Y}_{-\sigma,\sigma}(i\omega_l) \rbrace \,
  \Bar{\cal G}_2^{(0) \, ir}
  (\sigma,i\omega;\sigma,i\omega_l|\sigma,i\omega_l;\sigma,i\omega)
\nonumber \\ &-
  \frac{1}{\beta N} \sum_{{\bf k},\omega_l}
  \frac{\varepsilon^2({\bf k})}
       {D_{\sigma}({\bf k},i\omega_l)} \lbrace
  \Lambda_{-\sigma}(-i\omega_l) [
  1 - \varepsilon({\bf k}) \, \Lambda_{\sigma}(i\omega_l) ]
\nonumber \\ &-
  \varepsilon(-{\bf k}) \, Y_{\sigma,-\sigma}(i\omega_l)
  \Bar{Y}_{-\sigma,\sigma}(i\omega_l) \rbrace \,
  \Bar{\cal G}_2^{(0) \, ir}
  (\sigma,i\omega;-\sigma,i\omega_l| -\sigma,i\omega_l;\sigma,i\omega)
%
\end{align}
%
with
%
\begin{equation}
%
D_{\sigma}({\bf k},i\omega)
  = [1 - \varepsilon({\bf k}) \, \Lambda_{\sigma}(i\omega)]
    [1 - \varepsilon(-{\bf k}) \, \Lambda_{-\sigma}(-i\omega)]
  + \varepsilon({\bf k}) \, \varepsilon(-{\bf k}) \,
    Y_{\sigma,-\sigma}(i\omega) \,
    \Bar{Y}_{-\sigma,\sigma}(i\omega) .
%
\end{equation}
%
The corresponding equation for $\Bar{Y}_{-\sigma,\sigma}(i\omega)$
can be obtained from the expression for $Y_{\sigma,-\sigma}(i\omega)$.

The last equations together with the equations for the one- and
two-particle Green's functions determine completely the properties
of the superconducting phase, provided it exists. In order to gain
further insight into the physics contained in (47) and (48) we
will linearize the equations in terms of the order parameter
$Y_{\sigma,-\sigma}(i\omega)$ which will determine the
critical temperature $T_c$. The resulting equation for the
order parameter is of the form
%
\begin{align}
%
Y_{\sigma,-\sigma}(i\omega) = &- \frac{1}{\beta N} \sum_{{\bf k},\omega_l}
  \frac{\varepsilon({\bf k}) \, \varepsilon(-{\bf k})
        Y_{\sigma,-\sigma}(i\omega_l)}
       {[1 - \varepsilon({\bf k}) \, \Lambda_{\sigma}(i\omega_l)]
        [1 - \varepsilon(-{\bf k}) \, \Lambda_{-\sigma}(-i\omega_l)]}
\nonumber \\*[0.2cm] &\times
  \Bar{\cal G}_2^{(0) \, ir}
  (\sigma,i\omega;-\sigma,-i\omega|\sigma,i\omega_l;-\sigma,-i\omega_l) .
%
\end{align}
%
This equation must be solved together with the equation for
$\Lambda_{\sigma}(i\omega)$ which may be approximated by setting the
order parameters to zero giving,
%
\begin{align}
%
\Lambda_{\sigma}(i\omega) = {\cal G}_{p\sigma}^{(0)}(i\omega)
   &- \frac{1}{\beta N} \sum_{{\bf k},\omega_l}
      \frac{\varepsilon^2({\bf k}) \, \Lambda_{\sigma}(i\omega_l)}
           {1 - \varepsilon({\bf k}) \, \Lambda_{\sigma}(i\omega_l)}
   \, \Bar{\cal G}_2^{(0) \, ir}
   (\sigma,i\omega;\sigma,i\omega_l|\sigma,i\omega_l;\sigma,i\omega)
\nonumber \\ &-
   \frac{1}{\beta N} \sum_{{\bf k},\omega_l}
   \frac{\varepsilon^2({\bf k}) \, \Lambda_{-\sigma}(i\omega_l)}
           {1 - \varepsilon(-{\bf k}) \, \Lambda_{-\sigma}(i\omega_l)}
   \, \Bar{\cal G}_2^{(0) \, ir}
   (\sigma,i\omega;-\sigma,i\omega_l| -\sigma,i\omega_l;\sigma,i\omega) .
%
\end{align}
%

In order to determine $T_c$ we must solve (51) for $\Lambda_{\sigma}$
and insert the result in (50). The irreducible functions in (50) and
(51) can be written as
%
\begin{align}
%
\Bar{\cal G}_2^{(0) \, ir}
  &(\sigma,i\omega;\sigma,i\omega_l|\sigma,i\omega_l;\sigma,i\omega)
= \frac{(\omega - \omega_l)^2}{\Delta^2 \Delta_l^2}
  \Big\lbrace 2\omega_c^2(x + x_l) \tanh(\beta\epsilon/2)
\nonumber \\*[0.1cm] &-
  \frac{\coth(\beta\omega_c/2)}
       {\omega_c[(i\omega - i\omega_l)^2 - 4\omega_c^2]}
\left [ (xx_l + \omega_c^2)(\Delta\Delta_l + 8 \omega_c^4)
- 2\omega_c^2(\Delta + \Delta_l)(xx_l - \omega_c^2) \right ]
   \Big\rbrace ,
%
\end{align}
%
\begin{align}
%
\Bar{\cal G}_2^{(0) \, ir}
  &(\sigma,i\omega;-\sigma,i\omega_l|-\sigma,i\omega_l;\sigma,i\omega)
= - \frac{2\omega_c}{\Delta^2 \Delta_l^2}
    \Big \lbrace
    \omega_c(x + x_l)(\Delta + \Delta_l) \tanh(\beta\epsilon/2)
\nonumber  \\*[0.1cm] &+
    \coth(\beta\omega_c/2) (x + x_l)^2(xx_l - \omega_c^2) \Big \rbrace
   + \frac{\coth(\beta\omega_c/2) ((xx_l + 3\omega_c^2)}
         {\omega_c \Delta \Delta_l} ,
%
\end{align}
%
\begin{align}
%
\Bar{\cal G}_2^{(0) \, ir}
  &(\sigma,i\omega;-\sigma,-i\omega|\sigma,i\omega_l;-\sigma,-i\omega_l)
  =
\nonumber \\ &-
   \frac{2\epsilon(2\omega_c)^2 \tanh(\beta\epsilon/2)
          [i\omega \, i\omega_l + \epsilon^2 - \omega_c^2]}
        {[\omega^2 + (\epsilon + \omega_c)^2]
         [\omega^2 + (\epsilon - \omega_c)^2]
         [\omega_l^2 + (\epsilon + \omega_c)^2]
         [\omega_l^2 + (\epsilon - \omega_c)^2]}
\nonumber \\*[0.2cm] &+
   \frac{2\omega_c \coth(\beta\omega_c/2)
          [i\omega \, i\omega_l + 2 \omega_c
          (\epsilon - \omega_c)
        - (\epsilon - \omega_c)^2]}
        {[\omega^2 + (\epsilon - \omega_c)^2]
         [\omega_l^2 + (\epsilon - \omega_c)^2]
         [(\omega - \omega_l)^2 + (\epsilon + (2\omega_c)^2]}
\nonumber \\*[0.2cm] &+
   \frac{2\omega_c \coth(\beta\omega_c/2)
          [i\omega \, i\omega_l - 2 \omega_c
          (\epsilon + \omega_c)
        - (\epsilon + \omega_c)^2]}
        {[\omega^2 + (\epsilon + \omega_c)^2]
         [\omega_l^2 + (\epsilon + \omega_c)^2]
         [(\omega - \omega_l)^2 + (\epsilon + (2\omega_c)^2]} ,
%
\end{align}
%
where
%
\begin{subequations} \begin{align}
%
x &= i\omega - \epsilon, & \Delta &= (i\omega - \epsilon)^2 - \omega_c^2
 \\
x_l &= i\omega_l - \epsilon,
                & \Delta_l &= (i\omega_l - \epsilon)^2 - \omega_c^2 .
%
\end{align} \end{subequations}
%
For further discussion of (50) and (51) we introduce the
following shorthand notation:
%
\begin{align}
%
\phi_{\sigma}(i\omega) &=
   \frac{1}{N} \sum_{\bf k}
   \frac{\varepsilon^2({\bf k}) \, \Lambda_{\sigma}(i\omega)}
        {1 - \varepsilon({\bf k}) \, \Lambda_{\sigma}(i\omega)}
\nonumber \\ &=
   \frac{1}{N} \sum_{\bf k}
   \frac{\varepsilon({\bf k})}
        {1 - \varepsilon({\bf k}) \, \Lambda_{\sigma}(i\omega)} ,
\\
g_{\sigma}(i\omega)
 &= {\cal G}({\bf x} = {\bf x}^{\prime}|i\omega)
\nonumber \\ &=
   \frac{1}{N} \sum_{\bf k}
   \frac{\Lambda_{\sigma}(i\omega)}
        {1 - \varepsilon({\bf k}) \, \Lambda_{\sigma}(i\omega)} ,
\\
\phi_{\sigma}^{sc}(i\omega) &=
   \frac{1}{N} \sum_{\bf k}
   \frac{\varepsilon({\bf k}) \, \varepsilon(-{\bf k})}
    {[1 - \varepsilon({\bf k}) \, \Lambda_{\sigma}(i\omega)]
     [1 - \varepsilon(-{\bf k}) \, \Lambda_{-\sigma}(-i\omega)]}
\nonumber \\ &=
   \frac{\phi_{\sigma}(i\omega) - \phi_{-\sigma}(-i\omega)}
    {\Lambda_{\sigma}(i\omega) - \Lambda_{-\sigma}(-i\omega)} .
%
\end{align}
%
Furthermore we assume that
$\varepsilon({\bf k}) = \varepsilon(-{\bf k})$ holds with
$\sum_{\bf k} \varepsilon({\bf k}) =
\sum_{\bf k} \varepsilon^3({\bf k}) = 0$.
Sums will be replaced by integrals,
%
\begin{align}
%
\frac{1}{N} \sum_{\bf k} & = \int \! d\varepsilon \,
    \rho_0(\varepsilon) ,
\\
\rho_0(\varepsilon) &= \frac{4}{\pi W}
    \sqrt{1 - \left( \frac{2\varepsilon}{W}\right)^2} \times
    \begin{cases}
       1&     |\varepsilon| < \frac{W}{2} , \\
       0 &    |\varepsilon| < \frac{W}{2} ,
    \end{cases}
%
\end{align}
%
where $W$ is the band width and $\rho_0$ a model density of states
of semielliptic form. Since we do not consider magnetic states here,
the spin index in the paramagnetic phase can be omitted:
%
\begin{subequations}\begin{align}
%
\Lambda_{\sigma}(i\omega) &= \Lambda_{-\sigma}(i\omega)
                           = \Lambda(i\omega),
\\
\phi_{\sigma}(i\omega) &= \phi_{-\sigma}(i\omega)
                        = \phi(i\omega),
\\
\phi_{\sigma}^{sc}(i\omega) &= \phi_{-\sigma}^{sc}(i\omega)
                             = \phi^{sc}(i\omega) .
%
\end{align}\end{subequations}
%
However, the spin index is essential for the superconducting
order parameter $Y_{\sigma,-\sigma}(i\omega)$,
%
\begin{subequations}\begin{align}
%
Y_{\sigma,-\sigma}(i\omega) &= g_{\sigma,-\sigma}
     Y(i\omega),
\\
g_{\sigma,-\sigma} &= \delta_{\sigma,\uparrow}
                    - \delta_{\sigma,\downarrow} ,
%
\end{align}\end{subequations}
%
where $Y(i\omega)$ is an even function of the frequency,
%
\begin{equation}
%
 Y(i\omega) = Y(-i\omega) .
%
\end{equation}
%
Finally we have to add the equation which determines the chemical
potential:
%
\begin{equation}
%
\frac{1}{\beta} \sum_{\omega_n} \sum_{\sigma}
   {\cal G}_{\sigma}({\bf x} = {\bf x}^{\prime}|i\omega) \,
   e^{i\omega_n 0^+}
= \frac{2}{\beta} \sum_{\omega_n}
   g_{\sigma}(i\omega) \, e^{i\omega_n 0^+}
= \frac{N_p}{N} .
%
\end{equation}
%
Here $N_p$ is the number of polarons and $N$ the number of lattice sites.
With (59) and (60) the functions (56) and (57) can be written as
%
\begin{align}
%
\phi(i\omega) &= \frac{W}{2}
   \frac{\left( 1 - \sqrt{1 - \lambda^2(i\omega)} \right)^2}
        {\lambda^3(i\omega)}
 = \frac{W}{2} \frac{\lambda(i\omega)}
           {\left( 1 + \sqrt{1 - \lambda^2(i\omega)} \right)^2}
\\
g(i\omega) &= \frac{4}{W}
   \frac{\left( 1 - \sqrt{1 - \lambda^2(i\omega)} \right)^2}
        {\lambda(i\omega)}
 = \frac{4}{W} \frac{\lambda(i\omega)}
           {\left( 1 + \sqrt{1 - \lambda^2(i\omega)} \right)^2} ,
%
\end{align}
%
with $\lambda(i\omega) = (W/2) \, \Lambda(i\omega)$.
In order to check whether the state to be determined from
Eq.\ (51) is metallic or dielectric one has to analyze the
renormalized density of states given by
%
\begin{equation}
%
\rho(E) = - \frac{1}{\pi} \operatorname{Im} \, g(E + i0^+)
        = - \frac{1}{\pi} \operatorname{Im}
            \left( \frac{1 - \sqrt{1 - \lambda^2(E + i0^+)}}
            {\lambda(E + i0^+)} \right),
%
\end{equation}
%
where $\lambda(E + i0^+)$ is the analytical continuation
of $\lambda(i\omega)$.

\section{Analytical solutions}

The expressions for $\Lambda(i\omega)$ and  $Y(i\omega)$
simplify by using the shorthand notations (56) and (57) and the
symmetry property (62):
%
\begin{align}
%
Y(i\omega) &= \frac{1}{\beta} \sum_{\omega_l}
   \phi^{sc} (i\omega_l) \, \Bar{\cal G}^{(0) \, ir}
   (\sigma,i\omega;-\sigma,-i\omega|
   \sigma,i\omega_l;-\sigma,-i\omega_l) Y(i\omega_l) ,
\\
\Lambda(i\omega) &= {\cal G}_p^{(0)}(i\omega)
 - \frac{1}{\beta} \sum_{\omega_l}
   \phi(i\omega_l) \,\left [ \Bar{\cal G}^{(0) \, ir}
   (\sigma,i\omega;\sigma,i\omega_l|
   \sigma,i\omega_l;\sigma,i\omega) \right .
\nonumber \\
 & + \left . \Bar{\cal G}^{(0) \, ir}
   (\sigma,i\omega;-\sigma,i\omega_l|
   -\sigma,i\omega_l;\sigma,i\omega) \right ] .
%
\end{align}
%
In order to find a solution of Eq.\ (68) we insert (54),
replace $Y(i\omega_n)$ by
%
\begin{align}
%
Y(i\omega_n) &= \phi^{sc}(z_0) \, \chi(i\omega_n) \, Y(z_0)
 \quad \text{with $z \simeq 0$},
\\
\chi(i\omega_n) &= \frac{1}{\beta} \sum_{\omega_l}
   \Bar{\cal G}^{(0) \, ir}
   (\sigma,i\omega_n;-\sigma,-i\omega_n|
   \sigma,i\omega_l;-\sigma,-i\omega_l)
\nonumber \\ = \; &
\frac{2\omega_c [ \omega_c - \epsilon \tanh(\beta\epsilon/2)
                  \cosh(\beta\omega_c/2) ]
    + \cosh(\beta\omega_c) (-\omega_c^2 + \epsilon^2 + \omega_n^2)
      (\cosh(\beta\omega_c) - 1)^{-1}}
     {[\omega_n^2 + (\omega_c + \epsilon)^2]
      [\omega_n^2 + (\omega_c + \epsilon)^2]} .
%
\end{align}
%
and make use of Poisson's sum rule,
%
\begin{equation}
%
\frac{1}{\beta} \sum_{\omega_n} f(i\omega_n)
  = - \frac{1}{2\pi i} \int_C \! dz \,
      \frac{f(z)}{e^{\beta z} +1} ,
%
\end{equation}
%
where $C$ denotes the usual counterclockwise contour of the
imaginary axis. We obtain then from Eq.\ (68) with help of
the analytically continued function $\chi(z)$ for $Z = Z_0 = 0$
an equation for the critical temperature $T_c$,
%
\begin{equation}
%
\chi(0|\epsilon) \, \phi^{sc}(0|\epsilon) = 1 ,
%
\end{equation}
%
\begin{equation}
%
\chi(0|\epsilon) = \left( 2\omega_c
  [\omega_c - \epsilon \tanh(\beta_c\epsilon/2) \coth(\beta_c\omega_c/2)]
+ \frac{(\epsilon^2 - \omega_c^2) \cosh(\beta_c\omega_c)}
       {\cosh(\beta_c\omega_c) -1}
  \right)
  \left(\omega_c^2 - \epsilon^2\right)^{-2} .
%
\end{equation}
%
This quantity is even in $\epsilon$ and therefore only the absolute
value of $\epsilon = \Bar{\epsilon}_0 - \Bar{\mu}$ determines
$k_BT_c = \beta_c^{-1}$. From (58) and (65) one can make a rough
guess for the quantity $\phi^{sc}(0)$:
%
\begin{equation}
%
\phi^{sc}(0) \approx \left( \frac{W}{4} \right)^2
             \frac{1}{\gamma^2}, \quad
\gamma = \frac{1}{2} \left(
        1 + \sqrt{1 - \lambda^2(0 + i\delta)} \right) ,
%
\end{equation}
%
where $\gamma$ has to satisfy:
$\gamma(-\epsilon) = \gamma(\epsilon)$. This quantity can be obtained
self-consistently from Eq.\ (64) for the chemical potential.
For simplicity we replace here
$[1 + \sqrt{1 - \lambda^2(0 + i\delta)}]$
by $2\gamma$. Then (64) can be written as
%
\begin{equation}
%
\frac{2}{\gamma} \frac{1}{\beta} \sum_{\omega_n}
  \Lambda(i\omega_n) \, e^{i\omega_n0^+} = \frac{N_p}{N}.
%
\end{equation}
%
Using (69) together with (52) and (53) we can express
$\Lambda(i\omega_n)$ as
%
\begin{align}
%
\Lambda(i\omega_n) = &\;
  \frac{(i\omega_n - \epsilon)A_1(\epsilon) + \omega_c
        B_1(\epsilon)}
       {(i\omega_n - \epsilon)^2 - \omega_c^2}
\nonumber \\ &+
  \frac{\omega_c^2[(i\omega_n - \epsilon)A_2(\epsilon)
     + \omega_c B_2(\epsilon)]}
       {[(i\omega_n - \epsilon)^2 - \omega_c^2]^2}
%
\end{align}
%
with unknown coefficients $A_i$ and $B_i$. They can be found from
Eq.\ (69) or more easily from the asymptotic behavior of
(77) for $|\omega_n| \to \infty$,
%
\begin{align}
%
\underset{|\omega_n| \to \infty}{\Lambda(i\omega_n)} =&\;
  \frac{A_1}{i\omega_n} + \frac{A_1\epsilon + \omega_cB_1}
       {(i\omega_n)^2}
+ \frac{A_1(\omega_c^2 + \epsilon^2) + 2\epsilon\omega_cB_1
      + \omega_c^2A_2}{(i\omega_n)^3}
\nonumber \\ &+
  \frac{A_1(\epsilon^3 + 3\epsilon\omega_c^2)
      + B_1(\omega_c^3 + 3 \epsilon^2\omega_c)
      + \omega_c^2(3\epsilon A_2 + \omega_cB_2)}
       {(i\omega_n)^4}
      + \ldots
%
\end{align}
%
If we compare this with the asymptotic behavior of the full
one-polaron Green's function (Appendix A) by invoking the methods
of moments together with the asymptotic behavior of $g(i\omega_n)$
in (66), we obtain
%
\begin{subequations}\begin{align}
%
A_1(\epsilon) &= 1,
\\
B_1(\epsilon)  &= -\frac{1}{\omega_c} [M_1 + \epsilon] ,
\\
A_2(\epsilon)  &= \frac{1}{\omega_c^2} \left [
    M_2 + 2\epsilon M_1 + \epsilon^2 -\omega_c^2
    - (W/4)^2 \right ] ,
\\
B_2(\epsilon) &= \frac{1}{\omega_c^3} \left [
    - M_3 - 3\epsilon M_2 + M_1 \left(
    \omega_c^2 - 3\epsilon^2 + 3 (W/4)^2 \right)
    + \epsilon\omega_c^2 - \epsilon^3 + 3\epsilon
    (W/4)^2 \right ] ,
\label{equationd}
%
\end{align}\end{subequations}
%
where $M_i$ is the $i$-th moment of the one-polaron Green's function.
The results in (A.5) for the moments in lowest order allow to
evaluate $A_i$ and $B_i$, see (A.7). $A_1 =1$ describes the
asymptotic freedom of the polarons. $B_1 = \tanh(\beta\epsilon/2)$
is identical with its value in the zero-order polaron Green's function
of (38). The two new quantities $A_2$ and $B_2$ are small quantities
being proportional to $\omega_0/\omega_c = 1/\alpha$

Inserting now (77) into the left-hand part of Eq.\ (76) and
performing the summation leads to
%
\begin{align}
%
\frac{1}{\beta} & \sum_{\omega_n}
     \Lambda(i\omega_n) \, e^{i\omega_n 0^+} =
                                           \Bar{n}(\epsilon)
\nonumber \\ &+
\frac{\tanh(\beta\epsilon/2) \,
      [\tanh(\beta\omega_c/2) - 1][1 - \tanh^2(\beta\epsilon/2)]}
     {2[1 - \tanh^2(\beta\omega_c/2) \tanh^2(\beta\epsilon/2)]}
\nonumber \\*[0.1cm] &+
\frac{B_2(\epsilon)}{4} \tanh(\beta\omega_c/2)
     \frac{1 - \tanh^2(\beta\epsilon/2)}
          {1 - \tanh^2(\beta\epsilon/2) \tanh^2(\beta\omega_c/2)}
\nonumber \\*[0.1cm] &-
\frac{\beta\omega_c}{16} \left [
    \frac{A_2(\epsilon) + B_2(\epsilon)}
         {\cosh^2[\beta(\omega_c + \epsilon)/2]}
  - \frac{A_2(\epsilon) - B_2(\epsilon)}
         {\cosh^2[\beta(\omega_c - \epsilon)/2]} \right ] ,
%
\end{align}
%
which, according to Eq.\ (76), is equal to $(\gamma/2)(N_p/N)$.
Because of the large collective frequency, $\beta\omega_c \gg 1$,
we may omit exponentially small quantities like
$\exp(-\beta\omega_c)$. Since we are interested in results for
electron numbers which are close to half filling ($\epsilon = 0$),
also $|\epsilon| \ll \omega_c$ holds. Furthermore we will neglect
contributions of the order $1/\alpha$. Then the equation for the
chemical potential (74) is simply
%
\begin{equation}
%
\Bar{n}(\epsilon) =  \gamma n_p/2, \quad n_p = N_p/N .
%
\end{equation}
%
If furthermore we use $\gamma = 1$ (free polarons) we obtain with
(75) for $\phi^{sc}(0)$ the value
 %
\begin{equation}
%
\phi^{sc}(0 = (W/4)^2
%
\end{equation}
%
which allows to write the equation for the critical temperature
$T_c$ in the following form
%
\begin{equation}
%
\epsilon^2 + \omega_c^2 - 2\epsilon\omega_c \tanh(\beta\epsilon/2)
    = (\omega_c^2 - \epsilon^2)^2(4/W)^2 .
%
\end{equation}
%

In this approximation $T_c$ depends only on the local parameters;
but we expect that close to half filling this should give an
impression which of the local quantities is most important
for superconductivity in the strong-coupling limit of the
Hubbard-Holstein model. For strictly half filling Eq.\ (83)
can only be fulfilled when $\omega_c = W/4$. This may
perhaps be an unphysical large value for the renormalized
quantity. It also shows that the specific limit $\Bar{U} = 0$
is probably the critical value for the occurence of superconductivity
in the frame of the Hubbard-Holstein model. It is clear that for
$\Bar{U} < 0$ superconductivity is possible; but in this case it would
have to compete in energy with the energies of charge-ordered states.

For the special case $\omega_c = W/4$ we obtain
%
\begin{equation}
%
\left( \epsilon/\omega_c \right)^2
  \left [ 3 - \left( \epsilon/\omega_c \right)^2 \right ]
= \left( 2|\epsilon|/\omega_c \right) \tanh(\beta_c|\epsilon|/2) .
%
\end{equation}
%
Since $(\epsilon/\omega_c < 3$ holds
(not discussed in detail) we may seek for solutions for the case
$|\epsilon| \ll \omega_c$ leading to
%
\begin{align}
%
k_BT_c &= \frac{\omega_c}{3} \left [ 1 - \frac{5}{12}
  \left( \frac{\epsilon}{\omega_c} \right)^2 + \ldots \right ]
\nonumber \\*[0.2cm] &=
   \frac{W}{12} \left [ 1 - \frac{20}{3}
   \left( \frac{\epsilon}{W} \right)^2 + \ldots \right ] .
%
\end{align}
%

In spite of the many approximations (which however are all
reasonable) used the result for $T_c$ is in so far remarkable as it
shows that the critical temperature depends on the band width
(corresponding to the largest cut-off energy of the model)
and not on the effective mass of the ions. for small deviations
from half filling $T_c$ decreases and does not depend on the sign
of $\epsilon$.

For different values of $\omega_c$,
%
\begin{equation}
%
\omega_c = W/4 - y ,
%
\end{equation}
%
with $y \neq 0$ there are only solutions for not half filling.
In this case Eq.\ (84) may in this case be written as
%
\begin{align}
%
\beta_c |\epsilon| &= \ln \frac{1 + \kappa}{1 - \kappa} ,
\\*[0.1cm]
\kappa &= \frac{\epsilon^2 + \omega_c^2 -(4/W)^2(\omega_c^2 -
               \epsilon^2)^2}
              {2|\epsilon|\omega_c} , \quad 0 < \kappa < 1.
%
\end{align}
%
The condition $\kappa < 1$ is equivalent to
%
\begin{equation}
%
|\epsilon| + \omega_c < W/4
%
\end{equation}
%
while $\kappa > 0$ is related to the parameter $y$,
%
\begin{align}
%
\omega_c < W/4, \; y > 0&: \quad
   (W/4 - \omega_c)^2 < \epsilon^2 < \epsilon_{\max}^2 ,
\\
\omega_c > W/4, \; y < 0&: \quad
   \epsilon_{min}^2 < \epsilon^2 < \epsilon_{max}^2 ,
\\
\epsilon_{max, \, min}^2 = \omega_c^2 &+ (1/2)(W/4)^2
                    \pm (W/8)\sqrt{(W/4)^2 + 8\omega_c^2} .
%
\end{align}
%
For small $y$ we can simplify (87) and (88),
%
\begin{align}
%
\kappa \simeq &\;\, 2/(W|\epsilon|) \,
  \lbrace
  \epsilon^2 [3 - \epsilon^2(4/W)^2] + y[W/2 -4\epsilon^2/W
  - \epsilon^4(4(W)^3]
\nonumber \\ &+
  y^2 [-3 + \epsilon^2(4/W)^2 - \epsilon^4(4/W)^4] ,
  \rbrace
%
\end{align}
%
with the following restrictions for $\epsilon$:
%
\begin{align}
%
y > 0&: \quad y^2 < \epsilon^2 < 3(W/4)^2 -(5/6)Wy + (29/27)y^2 ,
\\
y < 0&: \quad (W/6)|y| + (25/27)y^2 < \epsilon^2 < 3(W/4)^2 +
        (5/6)W|y| + (29/27)y^2 .
%
\end{align}
%
Large values for $T_c$ can be achieved for $\kappa \nleqslant 1$
and in the vicinity of half filling ($\epsilon \neq 0$):
%
\begin{equation}
%
k_BT_c \simeq \frac{W\delta}{12(\delta - 1)}, \quad
     \delta = \frac{6\epsilon^2}{W|y|} < 1 , \quad
     y = \frac{W}{4} - \omega_c ,
%
\end{equation}
%
but only for $y < 0$ and hence $\omega_c > W/4$.

\section{summary}

We have discussed the occurence of superconductivity for the
Hubbard-Holstein model in the strong-coupling limit ($\Bar{g} \gg 1$).
Strong coupling leads to a renormalization of the one-polaron Green's
function already in the local approximation. For $\Bar{g} \gg 1$
we find a collective mode for the phonon clouds estimated by
evaluating integrals in the Laplace approximation. Due to the
absorption and emission of this mode by the polarons, the
on-site energies of polarons are renormalized. Similarly the irreducible
two-particle Green's are renormalized. Allowing for the exchange
of polarons including their phonon clouds leads to a new
irreducible Green's function which has been used to study spin-singlet
pairing of polarons. Analytical results for the superconducting
phase have been obtained for the limiting case, for which the
local repulsion of polarons is exactly canceled by their attractive
interaction. The resulting equation for the critical temperature
has been obtained by assuming a large collective-mode frequency
and a nearly half-filled band case. The parameters which determine
$T_c$ are $\omega_c \; (\geq W/4)$, $\epsilon$ ($\epsilon = 0$
corresponds to half filling) and the band width $W$ with $T_c$.
In the strong-coupling limit we obtain a critical temperature which is
of the order of $W/12$.

\begin{acknowledgments}

This work was supported by the Heisenberg-Landau Program and by
the Program of the High Council for Science and Technical Development
of Moldova. It is a pleasure to acknowledge discussions  with
Prof.\ Plakida and Prof.\ Crisan. V.\ A.\ M.\ would like to
thank the Universities of Duisburg and Salerno for financial
support. P.\ E.\ would like to thank the Bogoliubov Laboratory of
Theoretical Physics, JINR, for the hospitality he received
during his stay in Dubna.

\end{acknowledgments}

\appendix* \section{Method of moments}

By using the Heisenberg representation of the one-polaron Green'
function
%
\begin{align}
%
{\cal G}_{\sigma}({\bf x} - {\bf x}^{\prime}|\tau - \tau^{\prime})
  &= \langle T \,
    \hat{c}_{{\bf x}\sigma}(\tau) \,
    \hat{\Bar{c}}_{{\bf x}^{\prime}\sigma^{\prime}}(\tau^{\prime})
    \rangle_H ,
%
\end{align}
%
with
%
\begin{subequations}\begin{align}
%
\hat{c}_{{\bf x}\sigma}(\tau) &=
   e^{{\cal H}\tau} c_{{\bf x}\sigma} \, e^{-{\cal H}\tau} ,
\\
\hat{\Bar{c}}_{{\bf x}\sigma}(\tau) &=
   e^{{\cal H}\tau} c_{{\bf x}\sigma}^{\dagger} \, e^{-{\cal H}\tau} ,
%
\end{align}\end{subequations}
%
we can write the asimptotic expansion of the Fourier representation (25) for
$|\omega_n| \to \infty$ as
%
\begin{align}
%
{\cal G}_{\sigma}({\bf x} = 0|i\omega_n)
  &= g_{\sigma}(i\omega_n)
   = \frac{1}{i\omega_n} - \frac{M_1}{(i\omega_n)^2}
   + \frac{M_2}{(i\omega_n)^3}
   - \frac{M_3}{(i\omega_n)^4} + \ldots ,
\\*[0.2cm]
M_n &= \langle \lbrace c_{{\bf x}\sigma}^{\dagger}, \,
       \underbrace{
       [{\cal H} [{\cal H} \ldots [{\cal H}, \, c_{{\bf x}\sigma} ]
       }_{n}       \ldots ... ] \rbrace \rangle_H .
%
\end{align}
%
Here the statistical average $\langle \ldots \rangle_H$ is defined
with respect to the full density matrix of the grand canonical
ensemble. In the simplest approximation we obtain for the first three
moments of the Green's functions the relations,
%
\begin{subequations}\begin{align}
%
M_1 &= - (\epsilon + \omega_c \tanh(\beta\epsilon/2) ,
\\
M_2 &= \epsilon^2 + \omega_c^2 + (W/4)^2
      + 2\epsilon\omega_c \tanh(\beta\epsilon/2)
      + \omega_0 \omega_c \coth(\beta\omega_c/2) ,
\\
M_3 &= - \lbrace \epsilon^3 + 3\epsilon [ \omega_c^2
      + \omega_c\omega_0 \coth(\beta\omega_c/2) + (W/4)^2 ]
\nonumber \\ &+
        \omega_c \tanh(\beta\epsilon/2) \, [ \, 3 \epsilon^2
      + 3(W/4)^2 + \omega_c^2 + \omega_0^2
      + 3 \omega_0\omega_c \coth(\beta\omega_0/2) \, ] \rbrace .
%
\end{align}\end{subequations}
%
The expressions for the moments can be used to determine the
unknown coefficients, $A_n(\epsilon) \; A_n(\epsilon)$, in the
relation for $\Lambda_{\sigma}(i\omega)$, see Eq.\ (78),
by considering also the asymptotic behavior of $g_{\sigma}(i\omega)$
in (66) for small values of $\lambda_{\sigma}(i\omega)$,
%
\begin{equation}
%
g_{\sigma}(i\omega) = (2/W)\lambda_{\sigma}(i\omega) \, [ \,
   1 + (\lambda^2/4) + 2(\lambda^2/4)^2 + \ldots \; ] .
%
\end{equation}
%
We then insert the asymptotic form for $\lambda_{\sigma}(i\omega)$
from (78). Comparing the corresponding equations fixes the
coefficients $A_n(\epsilon)$ and $A_n(\epsilon)$,
%
\begin{subequations}\begin{align}
%
A_1(\epsilon) &= 1,
\\
B_1(\epsilon)  &= -\frac{1}{\omega_c} [M_1 + \epsilon] ,
      \simeq \tanh(\beta\epsilon/2) ,
\\
A_2(\epsilon)  &= \frac{1}{\omega_c^2} \left [
    M_2 + 2\epsilon M_1 + \epsilon^2 -\omega_c^2
    - (W/4)^2 \right ]
    \simeq  \frac{\omega_0}{\omega_c} \coth(\beta\omega/2) ,
\\
B_2(\epsilon) &= \frac{1}{\omega_c^3} \left [
    - M_3 - 3\epsilon M_2 + M_1 \left(
    \omega_c^2 - 3\epsilon^2 + 3 (W/4)^2 \right)
    + \epsilon\omega_c^2 - \epsilon^3 + 3\epsilon
    (W/4)^2 \right ]
\nonumber \\ &\simeq
     \frac{\omega_0}{\omega_c} \tanh(\beta\epsilon/2)
     \left [ \frac{\omega_0}{\omega_c} + 3 \coth(\beta\omega_0/2)
     \right ] .
%
\end{align}\end{subequations}
%



\end{document}